\documentclass[journal=nalefd,manuscript=article,layout=traditional]{achemso}

\usepackage{chemformula} 
\usepackage[T1]{fontenc} 
\usepackage{xcolor}
\usepackage{graphicx}
\usepackage{stfloats}

\author{J. Thoppil S}
\affiliation{School of Physical Sciences, Indian Institute of Technology Goa, Ponda, 403401, Goa, India}
\author{Y. Waheed}
\affiliation{School of Physical Sciences, Indian Institute of Technology Goa, Ponda, 403401, Goa, India}
\author{S. Shit}
\affiliation{School of Physical Sciences, Indian Institute of Technology Goa, Ponda, 403401, Goa, India}
\author{I. D. Prasad}
\affiliation{School of Physical Sciences, Indian Institute of Technology Goa, Ponda, 403401, Goa, India}
\author{K. Watanabe}
\affiliation{Research Center for Electronic and Optical Materials, National Institute for Materials Science,1-1 Namiki, Tsukuba 305-0044, Japan}
\author{T. Taniguchi}
\affiliation{International Center for Materials Nanoarchitectonics, National Institute for Materials Science, 1-1 Namiki, Tsukuba, 305-0044, Japan}
\author{S. Kumar}
\affiliation{School of Physical Sciences, Indian Institute of Technology Goa, Ponda, 403401, Goa, India}
\email{skumar@iitgoa.ac.in}
\phone{+91(0)-832-2490-138}

\title[An \textsf{achemso} demo]
  {Nanoparticle Stressor-Induced Single-photon Sources in Monolayer WS$_2$ Emitting into a Narrowband Visible Spectral Range}

\begin{document}

\begin{abstract}
A van der Waals heterostructure containing an atomically thin monolayer transition-metal dichalcogenide as a single-photon emitting layer is emerging as an intriguing solid-state quantum-photonic platform. Here, we report the utilization of spin-coating of silica nanoparticles for deterministically creating the spectrally isolated, energetically stable, and narrow-linewidth single-photon emitters in ML-WS$_2$. We also demonstrate that long-duration low-temperature annealing of the photonic heterostructure in the vacuum removes the energetically unstable emitters that are present due to fabrication-associated residue and lead to the emission of single-photons in a <\,25\,nm narrowband visible spectral range centered at $\sim$\,620\,nm. This work may pave the way toward realizing a hybrid-quantum-photonic platform containing a van der Waals heterostructure/device and an atomic-vapor system emitting/absorbing in the same visible spectral range.
\end{abstract}

Van der Waals heterostructures and devices of transition metal dichalcogenides (TMDs), hexagonal boron nitride (hBN), and graphene in their monolayer (ML) or a few-layer (FL) thicknesses regime have recently attracted significant attention of the scientific community working on exploitation of the fundamental and technological aspects of the quantum materials\cite{geim_van_2013,Manzeli2017}. Unlike their bulk counterparts, ML- and FL- TMDs possess intriguing physical properties such as reduced dielectric screening, large exciton binding energy, tunable bandgap, strong spin-orbit coupling, and spin-valley physics\cite{Manzeli2017}. Moreover, the ease with which the ML- and FL- TMDs can be integrated into photonic circuits and their abilities of hosting the single-photon emitters  (SPEs) have made the layered TMDs suitable candidates for various quantum technology applications.\cite{Jelena09QTech,aharonovich16QTech,gao23qkdwse2y,Montblanch2023,couteau23bio}. Although the physical origin of single-photon emission in TMDs is still under debate, several theoretical \cite{brooks18theory,parto2021theory,carmesin19theory,gupta18theory} and experimental \cite{tonndorf15,kumar15strain,branny17,palacios17,parto2021theory} studies have assigned it to an emission due to the spatially localized excitons. The formations of these excitons are driven either by local strain or the presence of optically active defects or due to the momentum-forbidden dark excitons or their combined effects\cite{lindlau18darkexciton}. Despite their microscopic origin, local-strain engineering is the most commonly used technique for deterministically creating the SPEs in layered TMDs.
The most commonly used method of local-strain engineering for creating position-controlled SPEs is based on the deposition/transfer of a ML-TMD layer onto a substrate that is pre-patterned with nanopillars utilizing an electron beam lithography technique\cite{kern16nanobar,branny17,palacios17}. The spatially-isolated SPE creation is achieved by introducing a local strain perturbation in the ML-TMD layer sitting at apex of the nanopillar through careful adjustment of the pillar height, aspect ratio, and pitch. Alternatively, nanoscale strain engineering has been achieved by placing the layered-TMDs on the nanostar structures\cite{peng20AuStar}. Other approaches of nanoscale strain engineering involve AFM-tip indentations of layered-TMDs via placing them onto either a flexible\cite{rosenberger19afm} or a flat\cite{xu21afm} substrate. The small sizes (10\,nm) metal-nanoparticles (NPs) coated with a dielectric material have also been used to create SPEs\cite{kim22PtNp} in 2D semiconductors.

While the majority of the studies/techniques mentioned above are focusing on SPEs in ML- or FL- WSe$_2$, some preliminary investigations have also been done for creating SPEs in ML- and bulk- WS$_2$ in the emission wavelength range of 600-730\,nm range. Although a photonic heterostructure containing ML-WS$_2$ has the potential to emit single photons in the visible spectral range where most commercially available detectors are highly efficient, the detailed investigations of such photonic heterostructure are yet to be explored. Recent studies have also highlighted that the SPEs in WS$_2$ may find their better usefulness for practical quantum key distribution technology\cite{vogl19radiation,Montblanch2023}. Although Palacios \textit{et al.}, \cite{palacios17} have shown the presence of SPEs at predefined locations, the emission lines of these SPEs exhibited complex comb-like structures on the shoulder of a defect-band emission. These comb-like spectral lines may pose various difficulties in further characterizing the optical properties of these SPEs. For example, even rigorous narrow-band spectral filtering would provide some photons due to the defect-band emission background, reducing the purity of the single-photon emission.

Here, we demonstrate a robust, cost-effective, and simple method for deterministically creating spatially- and spectrally- isolated SPEs in ML-WS$_2$. Our approach utilizes commercially available SiO$_2$ NPs to introduce local strain in the ML-WS$_2$ for creating the SPEs at deterministic locations. As-prepared samples/heterostructures were annealed in a vacuum and at a low temperature but for a longer duration to remove the long-wavelength and energetically-unstable SPEs, leading to the creation of energetically stable SPEs emitting into a $<$25\,nm (FWHM) narrowband visible spectral-range centered at $\sim$620\,nm.

\begin{figure*} [hbt] \centering
\includegraphics[width=18.5cm]{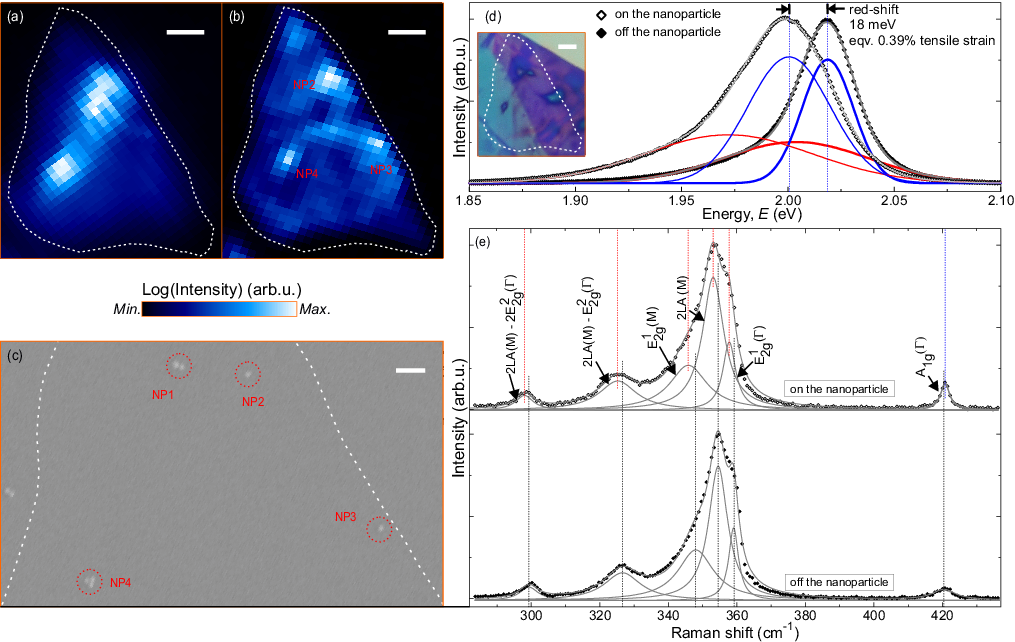}
\caption{Room-temperature \textmu-PL and \textmu-Raman spectroscopy measuring local strain due to a NP stressor. The spatial maps of integrated-intensity of PL in the spectral range of 575\,-\,750\,nm from (a) as-fabricated (b) annealed sample\,1. Scale bar: 2\,\textmu m. (c) SEM image of the sample\,1 after spin-coating of NPs on top of the bottom hBN. Red circles guide the presence of NPs, and white dotted lines indicate the WS$_2$ flake boundary. Scale bar: 1\,\textmu m. (d) \textmu -PL and (e) \textmu -Raman spectra from as-fabricated sample\,1 taken at-the-NP4 location (open diamonds) and off-the-NP location (closed diamonds). Solid thin- (thick-) lines are the fits for the spectrum at-the-NP4 (off-the-NP) location. Inset of (d): optical image of sample\,1 after annealing. Scale bar: 2\,\textmu m.}
\label{Fig:strain}
\end{figure*}

We started our investigations with an estimation of a local strain in ML-WS$_2$ caused by an NP stressor by performing the room-temperature micro-photoluminescence (\textmu-PL) and micro-Raman (\textmu-Raman) measurements, and the results are summarized in Figure\,\ref{Fig:strain}. The color-coded space maps of the integrated intensity of \textmu-PL emission from ML-WS$_2$ in the 575\,-\,750\,nm spectral range for as-fabricated and annealed sample\,1 are shown in Figure\,\ref{Fig:strain}a and Figure\,\ref{Fig:strain}b, respectively. The bright spots in both these PL space maps are due to enhancement in the intensity of excitonic emission owing to strain-induced funneling effect\cite{Moon2020}. Figure\,\ref{Fig:strain}c shows a scanning electron microscope (SEM) image of the bottom-hBN spin-coated with NPs. Although four distinct locations of NPs can be seen here, the PL space map of as-fabricated sample\,1 (Figure\,\ref{Fig:strain}a), shows only two bright spots due to the formation of a tent-like surface comprising NP1, NP2, and NP3. Remarkably, DI-water bathing followed by vacuum annealing has resulted in diffraction-limited emission spots (Figure\,\ref{Fig:strain}b) because of the conformality of ML-WS$_2$ with the NPs. An optical image of sample\,1 taken after annealing, as shown in the inset of Figure\,\ref{Fig:strain}d, shows this conformality.

The evolution of PL spectra of ML-WS$_2$ of as-fabricated sample\,1 at off- and on- the NP4 locations is shown in Figure\,\ref{Fig:strain}d; closed (open) diamonds are the measured data for off-the-NP4 (on-the-NP4) locations. We use two-lorenztian-peaks fitting (solid lines) method to extract emission energies of delocalized excitons. The thick- (thin-) solid lines are the fits for the PL spectrum taken at off-the-NP4 (on-the-NP4) locations. The high- (low-) energy peaks are assigned to the delocalized neutral-exciton, 2D-X$^0$, (single-negative-charged-exciton, 2D-X$^-$). The two vertical dotted lines indicate an 18\,meV redshift of the 2D-X$^0$ peak, which, as per literature report\cite{he16strainWS2} corresponds to a tensile strain of $\sim$0.39$\%$ at the NP4 location. The Raman spectroscopy measurements summarized in Figure\,\ref{Fig:strain}e further confirmed this strain value. The Raman spectrum from off- (on-) the-NP4 location is shown in the bottom (top) panel; the closed (open) diamonds are for off-(on-) the-NP4 location. The solid lines are the fits of a multilorenzian function. The vertical dotted lines indicate the peak positions of various Raman modes in ML-WS$_2$\cite{zeng13raman,zhang15raman,berk13raman}. The strain-dependent shifts of only a few Raman modes have been reported in the literature\cite{wang16strainRam}, whereas in our work, we see measurable changes in all the identified Raman modes. At the NP4 location, E$^1$$_{2g}$($\Gamma$) and 2LA(M) modes exhibit redshifts of 0.80  and 1.0 cm$^{-1}$, respectively, indicating the presence of a tensile strain of $\sim$0.39$\%$ which is in agreement with a strain measured using the PL emission.

\begin{figure*} [hbt] \centering
\includegraphics[width=15.2cm]{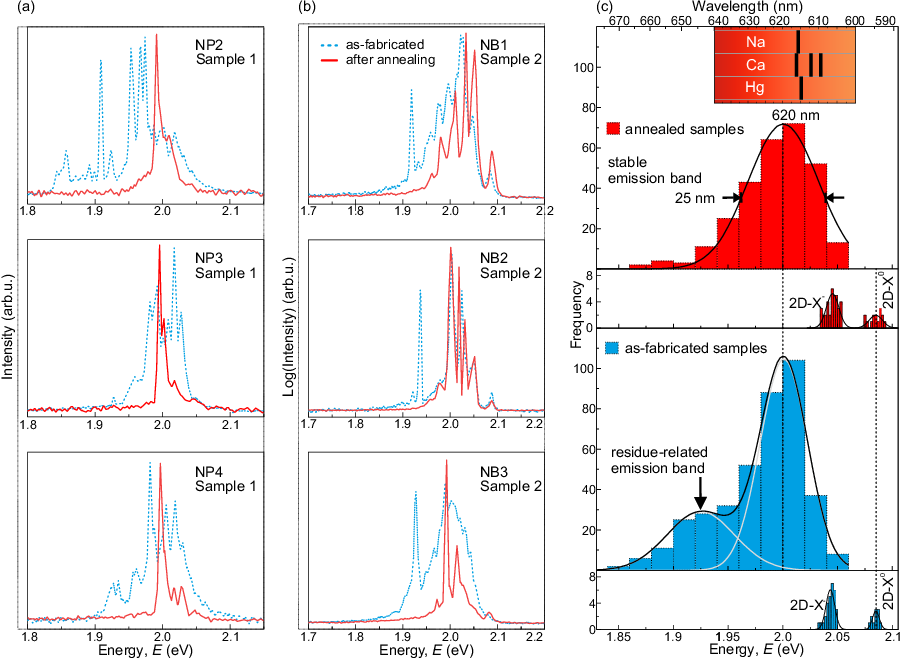}
\caption{Low-temperature vacuum annealing leading to the localized X$^0$ emission in a narrowband wavelength range. \textmu -PL spectra (a) from the NP locations of sample\,1 and (b) on NB locations of sample\,2. The spectra with dotted (solid) lines are for as-fabricated (annealed) samples. (c) Histograms of emission energies of localized exciton X$^0$  (bin size: 20\,meV), delocalized neutral exciton 2D-X$^0$ (bin size: 2\,meV) and delocalized negatively-charged exciton X$^-$ (bin size: 2\,meV) from as-fabricated (bottom-panel) and annealed samples (top-panel).}
\label{Fig:annealing}
\end{figure*}

We investigated the striking effects of low-temperature vacuum-annealing on the optical properties of SPEs originating from 5 samples with the same layered structures. The annealing treatments affected all five samples consistently in the same manner. Figure\,\ref{Fig:annealing} summarizes the \textmu-PL measurements at 4\,K for two samples. Figure\,\ref{Fig:annealing}a shows the PL emission spectra taken at the NP2, NP3, and NP4 locations of as-fabricated (dotted lines)and annealed (solid lines) sample\,1. A few nanometer-sized wrinkles, hereafter called nanobubbles (NBs), lead to the creation of localized SPEs\cite{zhang20}, and we also observed these NB-induced SPEs in sample\,2. Figure\,\ref{Fig:annealing}b shows the PL emission spectra taken at the NB1, NB2, and NB3 locations of as-fabricated (dotted lines) and annealed (solid lines) sample\,2. Both as-fabricated samples show isolated spectral lines of a few emitters from single-excitation spots, and their emission wavelengths are stretched in the 610\,-\,670\,nm spectral range. Figure\,\ref{Fig:annealing}c: bottom-panel shows a histogram of the emission wavelengths of all emitters originating from these two as-fabricated samples, and their emission can be categorized into two ensembles; a low-energy ensemble centered at 644\,nm and a high-energy ensemble centered at 620\,nm.

The comparisons of the emission spectra before annealing (dotted lines) and after annealing (solid lines) are shown in Figure\,\ref{Fig:annealing}a for sample\,1 and in Figure\,\ref{Fig:annealing}b for sample\,2. These demonstrate that the low-energy emission lines are absent in the annealed samples. The same is also evident by a single high-energy ensemble, centered at 620\,nm with an FWHM of 25\,nm, in the histogram of emission wavelengths of emitters from both the annealed samples shown in Figure\,\ref{Fig:annealing}c: top-panel. The ensemble peak position of the high-energy SPEs from both samples remains unchanged following the annealing treatment, which signifies that the annealing process has not altered the local-strain configuration, which is also evident from the histogram of the energies of 2D-X$^0$ and 2D-X$^{-}$ plotted in Figure\,\ref{Fig:annealing}c that their peak positions have not changed following the annealing treatment. The top panel of Figure\,\ref{Fig:annealing}c also marks the wavelengths of atomic transitions of Na, Ca, and Hg, and these wavelengths fall well within the range of the ensemble emission of SPEs from our samples. As the probability of finding two or more SPEs with identical emission wavelengths is very low, the post-processing techniques are utilized to tune the emission wavelength of one emitter to match the emission wavelength of other emitters. SPEs emitting into a narrowband of <\,25\,nm indicate that small electric- or magnetic- or strain-perturbative fields would be required to tune two or more emitters in resonances with each other or with other photonic platforms like solid-state or atomic vapor cells.

\begin{figure*} [hbt] \centering
\includegraphics [width=7.7cm]{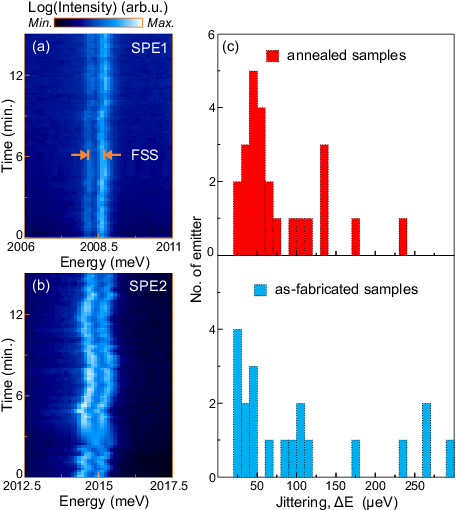}
\caption{Enhanced spectral stability and reduced linewidth due to low-temperature vacuum-annealing. Time trace of PL emission of a single emitter from ML-WS$_2$ (a) after and (b) before annealing, at $T$\,=\,4\,K. (c) Histogram of energy jitterings of various quantum emitters (bin size: 10 \textmu eV) before (bottom panel) and after (top panel) annealing.}
\label{Fig:jittering}
\end{figure*}

SPEs with energetically stable and narrow linewidth spectral lines are essential for a matured quantum photonic technology. Figure\,\ref{Fig:jittering} summarizes the effects of low-temperature vacuum-annealing on spectral-jittering (one-standard-deviation energy from the time-averaged emission energy, $\Delta E$) and linewidth (FWHM) of fine-structure split lines of SPEs. Figure\,\ref{Fig:jittering}a and Figure\,\ref{Fig:jittering}b present a 15 minutes-duration stack of PL spectra taken after every 10\,s for emitter SPE1 from annealed sample\,2 and emitter SPE2 from as-fabricated sample\,2, respectively. The doublet in both these color-coded intensity maps are the fine-structure split (FSS) lines of localized neutral excitons (X$^0$), and the slight changes in energies over time are due to the charge noise in the surrounding of these localized excitons.

The symmetry breaking of the confinement potential leads to the mixing of the two neutral excitonic states due to the electron-hole spin-exchange interaction, and therefore, the emission splits into two lines\cite{Gammon96,kumar15strain}. The lines of the doublet were fitted using two Gaussian functions. The measured $\Delta E$, of a SPE from the as-fabricated sample, as shown in Figure\,\ref{Fig:jittering}b, was 108 \textmu eV, and the measured linewidth was 356 \textmu eV. However, post-annealing, we observed significant improvements in both these parameters. The measured $\Delta E$ and linewidth, extracted for an SPE from the annealed sample\,2 as shown in Figure\,\ref{Fig:jittering}a, were reduced to 57 \textmu eV and 270 \textmu eV, respectively. The improvement in $\Delta E$, following the annealing treatment, is also evident from the statistical distributions of $\Delta E$ shown in Figure\,\ref{Fig:jittering}c: top-panel for emitters from the annealed samples, and in Figure\,\ref{Fig:jittering}c: bottom-panel for emitters from as-fabricated samples. For as-fabricated samples, we observed scattered values of $\Delta E$ ranging from 20\,-\,300 \textmu eV, whereas for the annealed samples, an ensemble-like $\Delta E$ distribution, centered at 45 \textmu eV, can be seen, and their values are ranging from 20\,-\,230 \textmu eV only signifying an improvement in the jittering mechanism.

We speculate that the physical origin of these low-energy and energetically unstable emitters is due to surface adsorbates, polymeric or organic impurities, or unstable crystal-defect states that wore off following the low-temperature vacuum-annealing treatment. Thus, the vanishing of these low-energy unstable emitters and improvement in the spectral jittering of the remaining high-energy emitters following the annealing process has turned out to be a crucial step for improving the optical quality of the emitters in the samples containing ML-WS$_2$. Two out of five samples have also shown a reduction in the emission intensities of these emitters, possibly due to the moisture trapped between the hBN and ML-WS$_2$ layers, which undergo the oxidation process at the annealing temperature.

\begin{figure*} [hbt] \centering
\includegraphics[width=17cm]{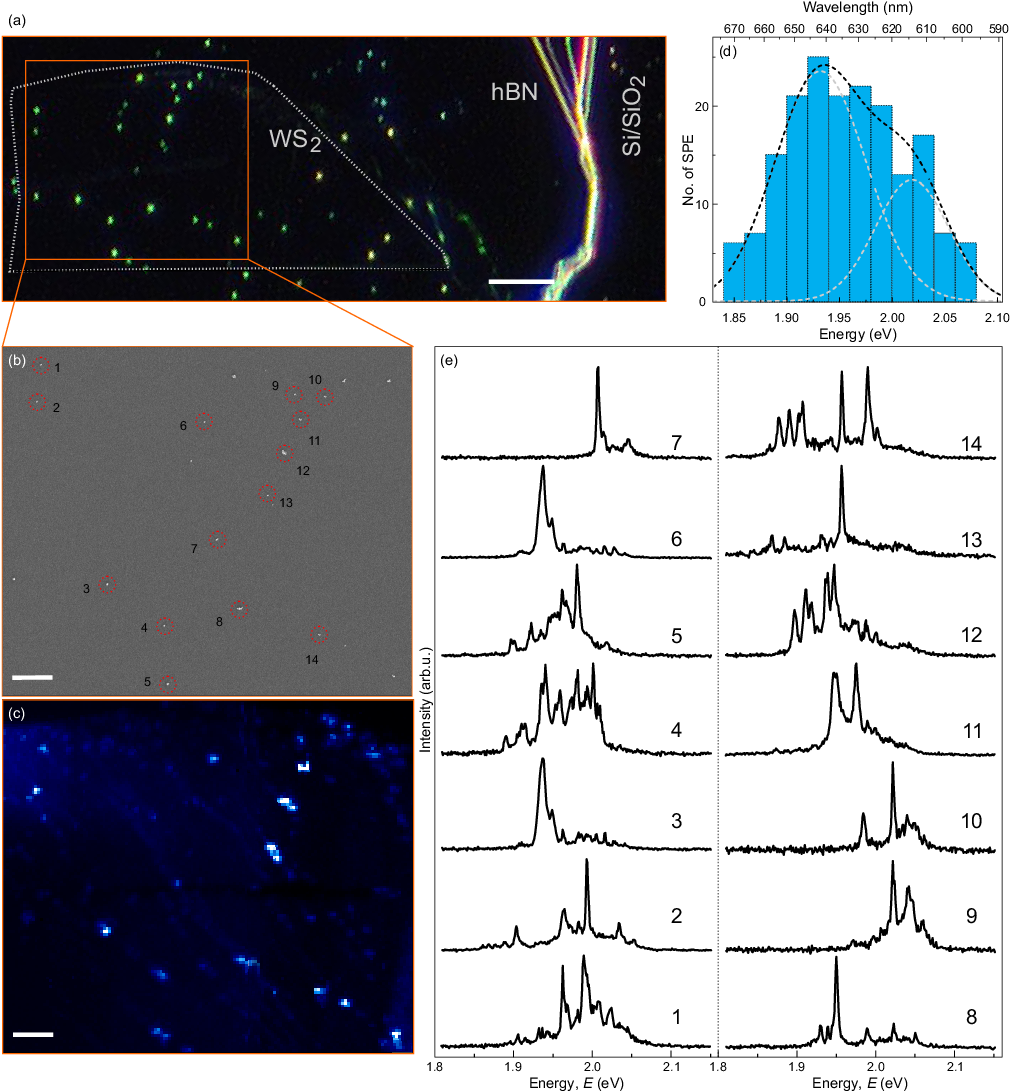}
\caption{Large scale realization of spatially localized SPEs. (a) Darkfield optical image of as-fabricated sample\,6. Scale bar: 20\,\textmu m. (b) SEM image of sample\,6 after spin-coating of NPs on top of the bottom h-BN. Red circles guide the presence of NPs. Scale bar: 5\,\textmu m. (c) Integrated intensity \textmu -PL spatial map in the 580\,-\,700\,nm spectral range at $T$\,=\,4\,K. Scale bar: 5\,\textmu m. (d) Histograms of emission energies of quantum emitters (bin size:\,20\,meV) from all NP locations on sample\,6. A bimodal Gaussian fit (red dotted lines) shows the presence of two ensembles of SPEs centered at 1.93 eV (642 nm) and 2.02 eV (614 nm). (f) PL spectra of ML-WS$_2$ acquired from NP sites marked in (b).}
\label{Fig:largescale}
\end{figure*}

\begin{figure*} [hbt] \centering
\includegraphics [width=16.0cm]{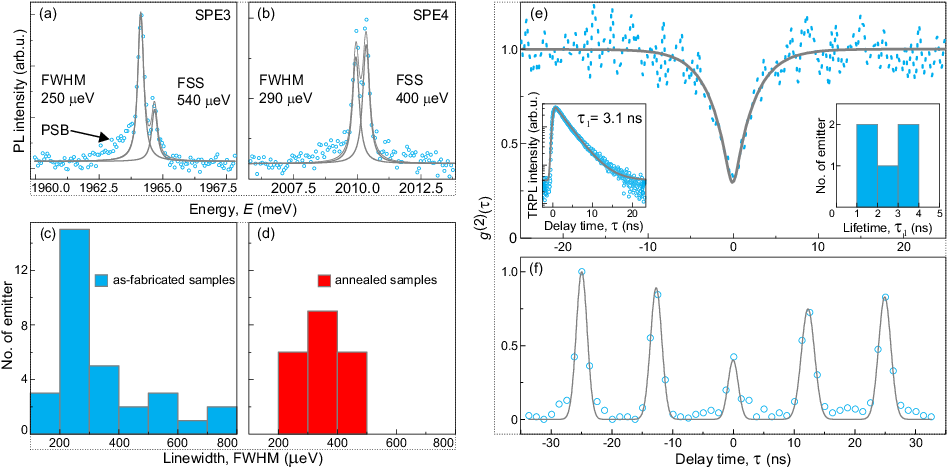}
\caption{High-resolution PL spectroscopy revealing background-free narrow linewidth spectral lines and photon correlations indicating on-demand single-photon emission. High-resolution PL X$^0$ emission spectra of emitter (a) SPE3 and (b) SPE4 from as-fabricated sample\,1 showing minimal photons emission into phonon sidebands. Open circles are the measurement data, and the solid lines are the Lorentzian fits used for extracting the linewidths and the fine-structure splittings. Histograms of linewidths (FWHMs) of various emitters from (c) as-fabricated and (d) annealed samples showing small inhomogeneous broadenings of emission lines. The bin sizes for (c) and (d) are 100\,\textmu eV. (e) Normalized second-order correlation function $g^{(2)}(\tau)$ as a function of delay time $\tau$ for X$^0$ lines of emitter SPE3 under the non-resonant CW excitation at 1\textmu W power. The dotted line is the measurement data, and the solid line is the fit yielding a deconvoluted $g^{(2)}(0)$\,=\,0.22 indicating a high-purity single-photon emission. (f) $g^{(2)}(\tau)$ for X$^0$ lines of emitter SPE4 under non-resonant pulsed-excitation at 520 nm wavelength indicating suppressed correlation counts at zero time delay indicating on-demand/turnstile single-photon emission. Insets of (e): Left: Time-Resolved PL intensity trace of emitter SPE3. Open circles are the measurement data, and the solid line is an exponentially modified Gaussian function fit showing a radiative lifetime of $\tau_1$\,=\,3.10\,ns. Right: A histogram of $\tau_1$ from various X$^0$ emitters showing a range of radiative lifetimes from 1\,-\,4\,ns.}
\label{Fig:antibunching}
\end{figure*}

Now, we show that the deterministic creation of SPEs using the spin-coating method is only limited by the sizes of the mechanically exfoliated ML-WS$_2$ flakes. All five heterostructures/samples discussed above contained small-sized flakes of ML-WS$_2$ because of several difficulties arriving during the fabrication processes. Figure\,\ref{Fig:largescale} demonstrates the creation of strain-induced, localized, and spectrally-isolated emitters with a high yield from an as-prepared sample\,6 containing large-size ML-WS$_2$ flake but without top-hBN encapsulation. Figure\,\ref{Fig:largescale}a shows a dark field optical image of the sample having an ML area of 125\,\textmu m by 62\,\textmu m. Figure\,\ref{Fig:largescale}b is an SEM image of NPs coated on bottom-hBN, which shows 22 NPs were present in an area of 2220\,\textmu m$^2$. Corresponding \textmu -PL map, as shown in Figure\,\ref{Fig:largescale}c, indicates ~\,90 \% yield in producing the strain-induced emitters, and the number of bright spots is only limited by the ML flake size. The emission energy histogram of these emitters, as shown in Figure\,\ref{Fig:largescale}d follows a bimodal distribution, with peaks centered at 642\,nm and 614\,nm, similar to as-prepared samples 1 and 2. Further complex comb-like spectral lines of 14 emitters from sample\,6, corresponding to 14 NP locations encircled in SEM image in Figure\,\ref{Fig:largescale}c, are shown in Figure\,\ref{Fig:largescale}e, and their spectral shapes are similar to the spectral shapes that have been observed for fully hBN-encapsulated samples 1 and 2.

Figure\,\ref{Fig:antibunching}a shows a high-resolution spectrum of emitter SPE3 from as-fabricated sample\,1, depicting the zero-phonon fine-structure split X$^0$ lines. A small fraction of emitted light goes into the phonon sideband (PSB) for emitter SPE3; however, we observed many more emitters from sample\,1 where negligible fractions of light go into the PSBs; the spectral shape of one such emitter SPE4 is shown in Figure\,\ref{Fig:antibunching}b. The doublet X$^0$ lines for SPE3 (SPE4) show an FSS of 540 (400) \textmu eV and a FWHM of 250 (290) \textmu eV. The linewidths of many more emitters, including those that do not show FSS, were measured from both samples. Figure\,\ref{Fig:antibunching}c and Figure\,\ref{Fig:antibunching}d show histograms of linewidths (FWHMs) of 31 emitters from as-fabricated and 21 emitters from annealed samples, respectively. Although we see a few emitters with nearly resolution-limited linewidths, the values of linewidths for these emitters are scattering in the range of 145\,-\,780\,\textmu eV. However, after annealing treatment, we see a visible narrowing in this range (260\,-\,460\,\textmu eV) with a well-defined distribution centered at 350\,\textmu eV demonstrating an overall improvement in the optical quality of most emitters.

Finally, we study the quantum-light nature of localized emitters. The photons-autocorrelation measurement on emitter SPE3 was performed using a standard HBT setup and under the non-resonant CW-excitation, as shown in Figure\,\ref{Fig:antibunching}e, which plots normalized second-order correlation function ($g^{(2)}(\tau)$), as a function of delay time ($\tau$). We obtain a deconvoluted $g^{(2)}(0)=0.22$, revealing the single-photon nature of the light\cite{Michler00} emitted by emitter SPE3. The characteristic time obtained from the second-order correlation measurement is 2.4\,ns. The measured X$^0$ radiative lifetime $\tau_1$ of the same emitter, obtained from the time-resolved PL measurement as shown in the inset (left) of Figure\,\ref{Fig:antibunching}e, is 3.1\,ns. A histogram of $\tau_1$ for five X$^0$ emitters is shown in inset (right) of Figure\,\ref{Fig:antibunching}e where we observed their values in 1\,-\,4\,ns range. A detailed investigation of $\tau_1$ for X$^0$ emission at $T\,=\,\sim$4\,K is unavailable in the literature. The values of $\tau_1$ for WSe$2$ emitters have been observed in a range of 1\,-\,10\,ns\cite{Cai18}, and we expect a similar range for emitters in WS$_2$ as well. The characteristic time obtained from $g^{(2)}(\tau)$ measurement for SPE3 is smaller than its $\tau_1$ value, indicating a fact that there might be other radiative or non-radiative channels through which X$^0$ may be coupling with the neighboring energy states\cite{Kumar16optica}. We further demonstrate an on-demand/turnstile single-photon emission by performing second-order correlation measurement on emitter SPE4 under non-resonant pulsed-excitation, as shown in Figure\,\ref{Fig:antibunching}f; where the value of $g^{(2)}(0)$ is 0.4 which is <\,0.5; a criterion of antibunched-light and on-demand/turnstile single-photons emission.

In summary, we demonstrated a simple, robust, and cost-effective method for deterministically creating strain-induced SPEs in ML WS$_2$ by utilizing commercially available SiO$_2$ NPs. Our sample fabrication method removed energetically unstable and low-energy emitters, leading to emission into a <25 nm narrowband centered at 620 nm with minimal spectral jittering <100\,\textmu eV. Due to a minimal inhomogeneous line broadening, we observed narrow linewidth emission lines from these emitters with an average linewidth (FWHM) of 350 \textmu eV. The emitters showed emission lines without a prominent defect-band shoulder in their emission spectra, which can reduce the single-photon purity. We showed a minimal to no fraction of emitted light for these emitters go into the PSB, one of the primary sources of decoherence mechanisms in solid-state quantum-photonic platforms.

Further, we demonstrated a 90\% yield in producing strain-induced SPEs, and the approach is limited only by the size of the ML flake and, therefore, can be extended for scalable quantum-photonic applications. Our method, combined with other nanoparticle-manipulator techniques, can create a periodically ordered set of high-quality SPEs on a chip for future linear-optical quantum technologies. Our work may pave the way towards realizing a hybrid-quantum-photonic platform containing a van der Waals heterostructure/device and an atomic-vapor system emitting/absorbing in the same visible spectral range.

\section*{Author Information}
\subsection*{Corresponding Author}
*E-mail (S.K.): skumar@iitgoa.ac.in
\subsection{Author Contributions}
J.T.S. and Y.W. contributed equally to this work.

\begin{acknowledgement}
We thank E.S. Kannan, S. R. Parne, and A. Rahman for the fruitful discussion and A. Rastelli for the data analysis software. This work was supported by the DST Nano Mission grant (DST/NM/TUE/QM-2/2019) and the matching grant from IIT Goa. I.D.P. thanks The Council of Scientific $\&$ Industrial Research (CSIR), New Delhi, for the doctoral fellowship. K.W. and T.T. acknowledge support from the JSPS KAKENHI (Grant Numbers 21H05233 and 23H02052) and World Premier International Research Center Initiative (WPI), MEXT, Japan.	
\end{acknowledgement}

\bibliography{achemsoarxiv}

\end{document}